\documentclass[5p,twocolumn,number,sort&compress]{elsarticle}

\usepackage[usenames]{color}
\usepackage{siunitx}
\usepackage{ulem}
\newcommand{\myvec}[1]{\mbox{\boldmath $#1$}}

\title{Simulating phase inversion processes by coupled map lattice:\\
Towards the theoretical design of food texture and quality in dairy processing from fresh cream to butter via whipped cream\vspace{-0.125cm}}
\date{\today}

\begin{document}

\author[1]{Erika Nozawa\corref{cor1}}
\ead{papers@e-rika.net}
\author[2]{Tetsuo Deguchi}
\cortext[cor1]{Corresponding author}
\address[1]{Graduate School of Organic Materials Science, Yamagata University, 4-3-16 Johnan, Yonezawa, Yamagata 992-8510, Japan}
\address[2]{Department of Physics, Faculty of Core Research, Ochanomizu University, 2-1-1 Ohtsuka, Bunkyo-ku, Tokyo 112-8610, Japan\vspace{-0.925cm}}
\begin{abstract}
We present a theoretical model and simulation for the formation dynamics of diverse texture patterns that emerge spontaneously or self-organize during phase inversion processes of fresh cream by mechanical whipping. The results suggest that the model should be applied for theoretically designing the texture and quality of whipped cream and butter products. The modeling complexity in phase inversion processes from fresh cream via whipped cream to butter was overcome by using a well-established complex systems approach, coupled map lattice (CML). The proposed CML consists of a minimal set of procedures (i.e., parameterized nonlinear maps), whipping, coalescence, and flocculation, acting on the appropriately coarse-grained field variables, surface energy, cohesive energy, and velocity (flow) of the emulsion defined on a two-dimensional square lattice. In the CML simulations, two well-known and different phase inversion processes are reproduced at high and low whipping temperatures. The overrun and viscosity changes simulated in these processes are at least qualitatively consistent with those observed in experiments. We characterize these processes exhibiting different texture patterns as the viscosity dominance at high whipping temperature and as the overrun dominance at low whipping temperature on the viscosity-overrun plane, which is one of the state diagrams.
\end{abstract}
\maketitle

\section{Introduction}
\label{introduction}

Fresh cream, a familiar and well-known dairy product, is a complex food consisting of a multi-component, multi-phase, and multi-scale emulsion that undergoes a phase inversion phenomenon to butter via whipped cream by the action of mechanical forces, i.e., whipping \cite{Fujita,Sato}.
During this phase inversion process, the extensive quasi-stable interfaces between water, fat globules, and air bubbles exhibit complex dynamic behavior \cite{Doi}, which causes diverse changes in the physical and chemical state and properties of fresh cream.
These changes, including aeration, flocculation, and coalescence \cite{Noda,Matsumura}, induce the spontaneous formation and self-organization of the spatial structure consisting of water, fat globules, air bubbles, and their interfaces.
This results in a series of quasi-stable spatial structures from fresh cream through whipped cream to butter, or diverse spatial textures (texture patterns) having different rheological properties \cite{Kaneda,Bourne}.

The texture and quality design of whipped cream and butter products necessitates a deep and complete understanding of the formation dynamics of the diverse texture patterns, including intermediate ones, that emerge during the phase inversion process.
While a theoretical simulation approach is considered essential to the understanding \cite{Masubuchi}, it has rarely been done to date due to the following circumstances.

Firstly, the phase inversion process occurs in an open complex system \cite{Kaneko}.
Therefore, it is necessary to formulate macroscopic equations that provide a consistent description of the emulsion state change under multi-combined elementary processes in phase inversion.
In such a description, high technical barriers for treating multi-component, multi-phase, and multi-scale emulsion make the combination of approximations complicated and the application of each approximation within its consistent range difficult.
Secondly, in order to elucidate the formation dynamics of texture patterns during the phase inversion process, it would be necessary to simulate, at a minimum, the state change of emulsion over a time span of approximately one millisecond and a size of approximately one millimeter squared.
Should this be attempted via microscopic simulation methods, such as molecular dynamics, a vast amount of calculations would be required ($10^{17}$ molecule numbers and $10^{12}$ time steps).
Such a calculation is currently infeasible.
In light of these circumstances, the breakthroughs with a complex systems approach to elucidating the formation dynamics of texture patterns have been eagerly anticipated \cite{Matsumura}.

The objective of this paper is to propose a coupled map lattice (CML) as a complex systems approach for modeling and simulating the formation of diverse texture patterns observed in phase inversion processes from fresh cream via whipped cream to butter.
As a dynamical system with discrete space and time and continuous state variables, CML is a powerful model class for the construction of dynamic and complex phenomena with the formation of diverse spatial patterns, through a series of nonlinear maps, called procedures \cite{Kaneko,Kaneko2}.
To date, numerous CMLs have been proposed, including those for boiling \cite{Yanagitab}, convection \cite{Yanagitac}, cloud formation \cite{Yanagitad}, sand ripple formation \cite{Nishimori}, and astronomical formation \cite{Nozawa,Nozawa2}.

This study presents a concise model construction with a minimal set of procedures that consistently describes multi-combined elementary processes in the phase inversion phenomenon.
The minimal procedure set is formulated with appropriately coarse-grained field variables of surface energy, cohesive energy, and velocity of the emulsion, as well as parameterized nonlinear maps of whipping, where the emulsion is aerated, coalescence, where the emulsion is changed from partial to complete in its coalesced state, and flocculation, where the emulsion is demulsified.
This construction not only circumvents the aforementioned approximation and computation challenges, but also enables us to recognize important elementary processes that are necessary (or essential) for reproducing the formation of texture patterns observed in phase inversion processes.
This is due to the primary characteristic of CML known as ``reductionism in procedure'' \cite{Kaneko}.

Furthermore, this study reproduces two different phase inversion processes at high and low whipping temperatures, which are well-known in the dairy processing and confectionery industries.
These processes exhibit different texture patterns, and are characterized as the viscosity dominance at high whipping temperature and as the overrun dominance at low whipping temperature on the viscosity-overrun plane.
The observed changes in overrun and viscosity for these processes are at least qualitatively consistent with the experimental results \cite{Matsumoto,Okuyama,Ihara} and confectionery techniques \cite{Kajiwara}.
This reproduction (or potential prediction) is fully executable on the personal computer level.
It is based on the secondary characteristic of CML, known as a global search ability which extends even into unknown regions of parameters controlling the procedures \cite{Yanagitab,Yanagitac,Yanagitad}.
The secondary characteristic is provided by the fast computation enabled by the primary characteristic of CML.

The present paper is organized as follows.
In Section \ref{model}, we construct a CML for the phase inversion phenomenon in accordance with the general prescription of CML construction: the introduction of a lattice, the assignment of field variables, and the formulation of procedures.
In Section \ref{simulation of two different phase inversion processes}, we present the simulation results of the two different phase inversion processes at high and low whipping temperatures, using a series of texture patterns.
In Section \ref{analysis of two different phase inversion processes}, we present a detailed analysis of these processes, based on the overrun and viscosity curves with respect to whipping time and the process curves on the viscosity-overrun plane.
We characterize these processes as the viscosity dominance at high whipping temperature and as the overrun dominance at low whipping temperature.
Summary and discussion are given in Section \ref{summary and discussion}.

\section{Model}
\label{model}

\begin{figure}[h]
\begin{center}
  \includegraphics[scale=0.5]{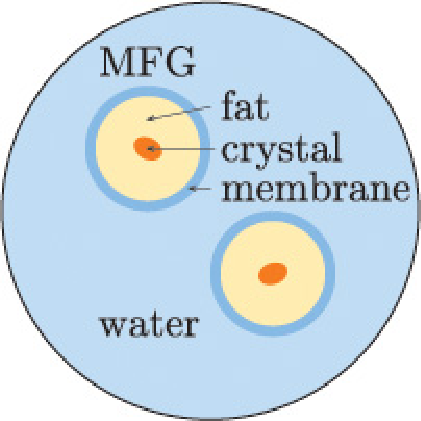}
\end{center}
\vspace{-0.5cm}
\caption{\footnotesize Schematic illustration of fresh cream consisting of water and milk fat globules (MFGs). The MFGs are further composed of fat droplets (yellow spheres), lipid crystals (orange ellipsoids), and milk fat globule membranes (blue spherical shells surrounding the fat droplets).}
\label{fig:emu.eps}
\vspace{-0.4cm}
\end{figure}

We construct a coupled map lattice (CML\cite{Kaneko,Kaneko2}) for simulating the phase inversion processes of fresh cream, which consists of water and milk fat globules (MFGs) as shown in Fig.\ref{fig:emu.eps}.
Hereafter, fresh cream (O/W emulsion), whipped cream (foam) and butter (W/O emulsion), including their intermediate states, are simply referred to as ``emulsions'' consisting of water, MFGs and air.
The following subsections present a step-by-step modeling of the dynamic behavior of fresh cream in the phase inversion process, according to the general prescription of CML construction \cite{Kaneko,Kaneko2}.

\subsection{Emulsion on the lattice}
\label{emulsion on the lattice}

The first construction step is to introduce coordinates to an object of interest using a suitable lattice. In our model, we consider an emulsion filling a relatively flat container. Viewed from above, this is represented approximately as a virtual emulsion on a two-dimensional square lattice.
The lattice points are denoted by $ij$ ($i=0,1,\cdots,N_{x}-1$ and $j=0,1,\cdots,N_{y}-1$) and their position vectors by $\myvec{r}_{ij}$ $=(i,j)=i\myvec{e}_{x}+j\myvec{e}_{y}$. Here, $\myvec{e}_{x}$ and $\myvec{e}_{y}$ are the unit vectors in the $x$- and $y$-axis directions, respectively. The emulsion at lattice point $ij$ is defined as a collection of virtual particles uniformly distributed in the square cell of size one centered at $ij$ (see the ``particle picture'' in Ref.\cite{Nozawa}).

\subsection{Field variables}
\label{field variables}

The second construction step is to assign field variables representing the state of the object at discrete time $t$ to the lattice. In our model, we consider typical types of emulsions in the phase inversion processes that will be discussed shortly, and introduce the surface energy $s_{ij}^{t}$, cohesive energy $c_{ij}^{t}$, and velocity (flow) $\myvec{v}_{ij}^{t}=v_{x\, ij}^{t}\myvec{e}_{x}+v_{y\, ij}^{t}\myvec{e}_{y}$ of the emulsion, and the emulsion energy $h_{ij}^{t}=s_{ij}^{t}+c_{ij}^{t}$ as physical and chemical field variables at lattice point $ij$.

Surface energy $s_{ij}^{t}$ and cohesive energy $c_{ij}^{t}$ are derived from the formation and destruction of interfaces in the emulsion, respectively. We consider various interfaces such as the water-fat globule, air-fat globule, and water-air interfaces. Using these field variables, three typical types of emulsions can be represented in a nutshell: (1) Fresh cream with low $s_{ij}^{t}$ and low $c_{ij}^{t}$; (2) Whipped cream with high $s_{ij}^{t}$ and low $c_{ij}^{t}$; (3) Butter with low $s_{ij}^{t}$ and high $c_{ij}^{t}$.

\subsection{Elementary processes and procedures}
\label{elementary processes and procedures}

The third construction step is to formulate procedures describing the elementary processes of state change of the object using relatively simple nonlinear maps acting on the field variables. In our model, we consider only important elementary processes of physical and chemical changes of the emulsion in the phase inversion phenomenon, and formulate the whipping procedure $T_{w}$, the coalescence procedure $T_{c}$, and the flocculation procedure $T_{f}$.

For the flexible and proper formulation, CML generally offers two types of procedures, the Eulerian and the Lagrangian procedures \cite{Kaneko,Yanagitac}. The Eulerian procedures describe the change in field variables on each lattice point by local and global interactions (see ``lattice picture'' in Ref.\cite{Nozawa}), and the Lagrangian procedures describe the change in field variables in each cell along the flow of the virtual particles (see ``particle picture'' in Ref.\cite{Nozawa}). In the following subsections, the whipping procedure $T_{w}$ is given as a Lagrangian procedure and the coalescence procedure $T_{c}$ and flocculation procedure $T_{f}$ are given as Eulerian procedures, respectively.

\subsubsection{Whipping procedure}
\label{whipping procedure}

\begin{figure}[h]
\begin{center}
  \includegraphics[scale=0.5]{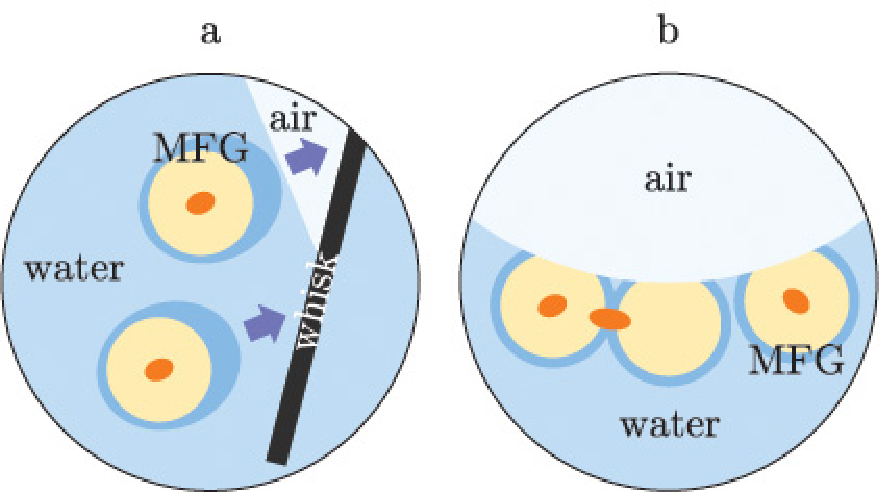}
\end{center}
\vspace{-0.5cm}
\caption{\footnotesize Conceptual illustration of the whipping procedure $T_{w}$. (a) Whipping-induced flow carries the emulsion of water and MFGs while deforming the water-fat globule interface (i.e., MFG membranes). (b) Deformation of membranes increases the surface energy of the emulsion and leads to aeration.}
\label{fig:tw.eps}
\end{figure}

In the whipping procedure $T_{w}$, whipping-induced flows carry the emulsion while deforming the water-fat globule interface (i.e., MFG membranes) as shown in Fig.\ref{fig:tw.eps}a. This membrane deformation that has occurred at the mesoscale leads to aeration \cite{Fujita} (in confectionery terms) at the macroscale where MFGs surround air bubbles as shown in Fig.\ref{fig:tw.eps}b, which increases the surface energy of the emulsion. In the following we describe, in turn, how the surface energy increases from the mesoscale to the macroscale.

When discrete whipping with angular speed $\omega$ is added at time interval $\iota$, the whipping velocity $\myvec{w}_{ij}^{t}$ at lattice point $ij$ becomes
\begin{equation}
\myvec{w}_{ij}^{t}=\left\{
\begin{array}{lr}
\omega\Bigl\{\Bigl(\frac{N_{y}-1}{2}-j\Bigr)\myvec{e}_{x}\Bigr.&\\
\hspace{0.2cm}\Bigl.+\Bigl( i-\frac{N_{x}-1}{2}\Bigr)\myvec{e}_{y}\Bigr\},&t\bmod\iota=0,\\
0,&\textrm{otherwise},
\end{array}
\right.
\end{equation}
where $((N_{x}-1)/2,(N_{y}-1)/2)$ is the center position of the lattice.
This whipping carries the emulsion at lattice point $ij$ to $\tilde{\myvec{r}}_{ij}^{t}$ while changing its velocity from $\myvec{v}_{ij}^{t}$ to $\myvec{w}_{ij}^{t}$ over a relaxation time $\tau$.
Here the velocity during relaxation $\myvec{v}(t')$ follows $d\myvec{v}/dt'=-(\myvec{v}-\myvec{w}_{ij}^{t})/\tau$.

Let us briefly explain the physical background of the whipping procedure.
The constituent particles of a complex emulsion such as water, MFGs, and air bubbles move together without phase separation in the emulsion flows, and hence they can be regarded as relatively simple clustered units (water- and air-clad MFGs or MFG-clad water and air bubbles as shown in Fig.\ref{fig:tw.eps}), even though they are of multi-component, multi-phase, and multi-scale.
This observation leads to a concise description for the complex motion of an emulsion, that is, a relaxation mechanism based on the difference in mass (i.e., relaxation time) of water and MFG particles, and a deformation mechanism of their interface (i.e., the MFG membrane in Fig.\ref{fig:tw.eps}a) that occurs due to this mechanism.

\begin{figure*}[t]
\begin{center}
	\includegraphics[scale=0.5]{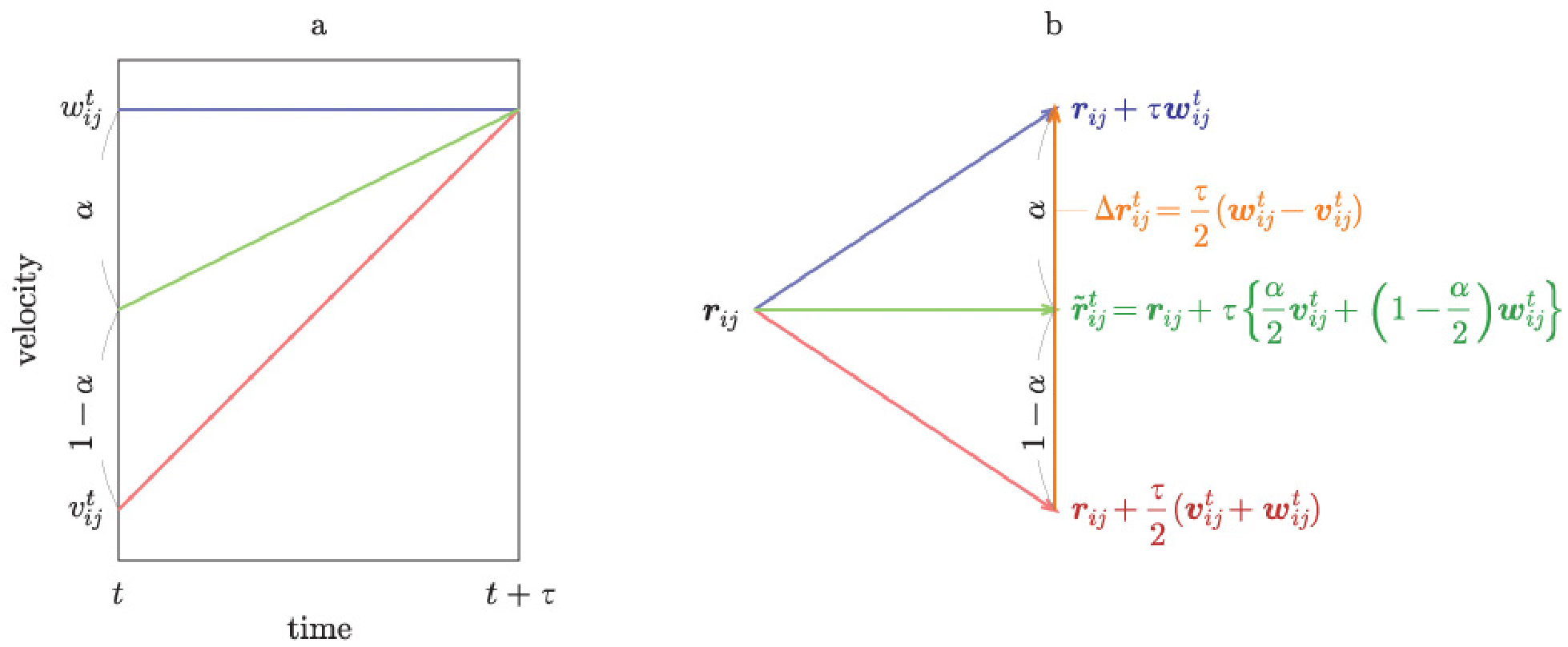}
\end{center}
\vspace{-0.5cm}
\caption{\footnotesize Difference in relaxation behavior with respect to flow between water and MFGs. (a) Velocity changes. The blue, red, and green lines represent the velocity changes of a water particle, MFG particle, and emulsion at lattice point $ij$, respectively. (b) Displacements. The blue, red, and green arrows represent the displacements of the water particle, MFG particle, and emulsion, respectively. The orange arrow represents the displacement of MFG membrane between the water and MFG particles. Here mixing coefficient $\alpha$ is 0.5.}
\label{fig:rel_cur.eps}
\end{figure*}

We now approximate the solution of the above velocity relaxation with a piecewise linear function $\myvec{v}(t')=\myvec{v}(t)-(\myvec{v}(t)-\myvec{w}_{ij}^{t})(t'-t)/\tau$ for $t\le t'<t+\tau$ and $\myvec{w}_{ij}^{t}$ for $t'\ge t+\tau$, as shown in Fig.\ref{fig:rel_cur.eps}.
The velocity changes of water and MFG particles during relaxation are shown in Fig.\ref{fig:rel_cur.eps}a, and their displacements in Fig.\ref{fig:rel_cur.eps}b.
Water particles are so light that they immediately relax into flow $\myvec{w}_{ij}^{t}$ (the blue line in Fig.\ref{fig:rel_cur.eps}a) and are thus carried to position $\myvec{r}_{ij}+\tau\myvec{w}_{ij}^{t}$ (the blue arrow in Fig.\ref{fig:rel_cur.eps}b). On the other hand, MFG particles are so heavy that they gradually relax into flow $\myvec{w}_{ij}^{t}$ over relaxation time $\tau$ (the red line in Fig.\ref{fig:rel_cur.eps}a) and are thus carried to position $\myvec{r}_{ij}+\frac{\tau}{2}(\myvec{v}_{ij}^{t}+\myvec{w}_{ij}^{t})$ (the red arrow in Fig.\ref{fig:rel_cur.eps}b). It thus follows that the emulsion is carried to position $\tilde{\myvec{r}}_{ij}^{t}=(\tilde{\imath}^{t},\tilde{\jmath}^{t})$ (the green arrow in Fig.\ref{fig:rel_cur.eps}b) given by,
\begin{eqnarray}
\lefteqn{
\tilde{\myvec{r}}_{ij}^{t}
=\tilde{\myvec{r}}\left(\myvec{r}_{ij},\myvec{w}_{ij}^{t},\myvec{v}_{ij}^{t}\right)
}
\hphantom{\hspace{7.4cm}}
\nonumber\\
\lefteqn{
=(1-\alpha)\left(\myvec{r}_{ij}+\tau\myvec{w}_{ij}^{t}\right)
+\alpha\left\{\myvec{r}_{ij}+\frac{\tau}{2}\left(\myvec{v}_{ij}^{t}+\myvec{w}_{ij}^{t}\right)\right\}
}
\hphantom{\hspace{7.4cm}}
\nonumber\\
\lefteqn{
=\myvec{r}_{ij}+\tau\left\{\frac{\alpha}{2}\myvec{v}_{ij}^{t}+\left(1-\frac{\alpha}{2}\right)\myvec{w}_{ij}^{t}\right\},
}
\hphantom{\hspace{7.4cm}}
\end{eqnarray}
with the velocity of the center of gravity of the emulsion (the green line in Fig.\ref{fig:rel_cur.eps}a).
Here $\alpha$ is the mixing coefficient representing the mass fraction of MFGs in the emulsion.

\begin{figure*}[t]
\begin{center}
	\includegraphics[scale=0.5]{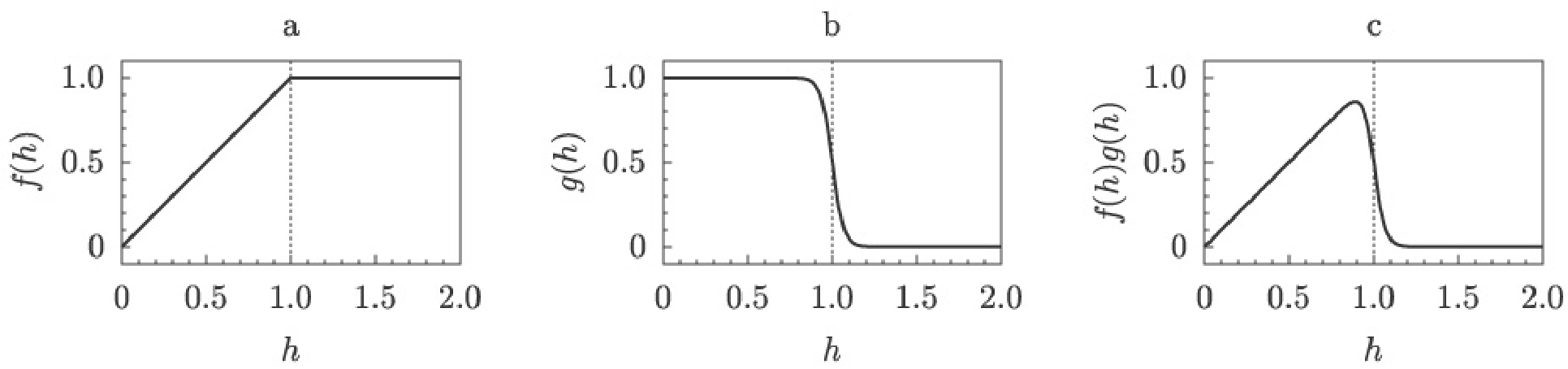}
\end{center}
\vspace{-0.5cm}
\caption{\footnotesize (a) Residual factor $f(h)$. (b) Occurrence factor $g(h)$. (c) Aeration factor $f(h)g(h)$. The horizontal axis is emulsion energy $h$. Here threshold $\theta$ for the surface activity of MFG membranes is 1.
}
\label{fig:f_g_fg.eps}
\end{figure*}

The difference in relaxation between water and MFGs generates the displacement of the water-fat globule interface (i.e., MFG membrane)
\begin{equation}
\Delta\myvec{r}_{ij}^{t}=\Delta\myvec{r}\left(\myvec{w}_{ij}^{t},\myvec{v}_{ij}^{t}\right)=\frac{\tau}{2}(\myvec{w}_{ij}^{t}-\myvec{v}_{ij}^{t}),
\end{equation}
shown as the orange arrow in Fig.\ref{fig:rel_cur.eps}b.
For simplicity, we assume that the surface energy increment $\Delta s_{ij}^{t}$ is given by
\begin{eqnarray}
\lefteqn{
\Delta s_{ij}^{t}
=\Delta s\left(\Delta\myvec{r}_{ij}^{t}\right)
}
\hphantom{\hspace{7.4cm}}
\nonumber\\
\lefteqn{
=\frac{1}{2}\alpha\kappa\left(\Delta\myvec{r}_{ij}\right)^{2}
=\frac{1}{8}\alpha\kappa\tau^{2}\left(\myvec{w}_{ij}^{t}-\myvec{v}_{ij}^{t}\right)^{2},
}
\hphantom{\hspace{7.4cm}}
\end{eqnarray}
where $\kappa$ is the elastic coefficient of MFG membranes.

Surface energy increment $\Delta s_{ij}^{t}$ due to membrane deformation at the mesoscale leads to aeration at the macroscale in the following way (Fig.\ref{fig:tw.eps}b).
MFGs adsorb to introduced air bubbles via proteins \cite{Noda,Brooker} in an attempt to quickly reduce the surface energy caused by the membrane deformation.
The adsorbing MFGs partially coalesce one after the other via their lipid crystals and cover the bubble surface \cite{Fujita,Sato}.
The MFG-covered bubbles cross-link with adjacent bubbles via MFGs to form a more rigid, quasi-stable structure \cite{Noda,Brooker}, and the emulsion is now aerated.
Thus, surface energy increment $\Delta s_{ij}^{t}$ in the unstable membrane deformation at the mesoscale remains to some extent as a surface energy increment in the quasi-stable aeration at the macroscale.

There are two important factors that affect the extent of aeration \cite{Fujita}. The first one is a factor representing the residual rate of surface energy increment $\Delta s_{ij}^{t}$. This gives a measure of the ease of aeration, and increases and saturates as the emulsion is aerated by the coalesced MFGs (i.e., surface energy $s_{ij}^{t}$ increases), and as the emulsion is flocculated by the introduced air bubbles (i.e., cohesive energy $c_{ij}^{t}$ increases). The second one is a factor representing the occurrence rate of surface energy increment $\Delta s_{ij}^{t}$. This gives a measure of the ease of membrane deformation, and decreases sharply when the multiplicative effect between aeration (an increase in surface energy $s_{ij}^{t}$) and flocculation (an increase in cohesive energy $c_{ij}^{t}$) reduces the surface activity of MFG membranes.

We express these two factors as two functions of emulsion energy $h_{ij}^{t}$ ($=s_{ij}^{t}+c_{ij}^{t}$), with a temperature-dependent threshold $\theta$ and gain coefficient $\beta$ for the surface activity of MFG membranes, based on Refs.\cite{Fujita,Dickinson,Walstra}. The residual factor $f(h_{ij}^{t})$ (Fig.\ref{fig:f_g_fg.eps}a) is given by
\begin{equation}
f(h_{ij}^{t})=\left\{
\begin{array}{ll}
\frac{h_{ij}^{t}}{\theta},&h_{ij}^{t}\le\theta,\\
1,&h_{ij}^{t}>\theta,
\end{array}
\right.
\end{equation}
and the occurrence factor $g(h_{ij}^{t})$ (Fig.\ref{fig:f_g_fg.eps}b) is given by
\begin{equation}
\label{eq:h(f)}
g(h_{ij}^{t})=\frac{1}{1+e^{\beta(h_{ij}^{t}-\theta)}}.
\end{equation}
Hence, the extent of aeration (i.e., the surface energy increment at the macroscale) is expressed as $f(h_{ij}^{t})g(h_{ij}^{t})\Delta s_{ij}^{t}$. From now, we call $f(h_{ij}^{t})g(h_{ij}^{t})$ the aeration factor (Fig.\ref{fig:f_g_fg.eps}c).

The whipping procedure $T_{w}$ is therefore defined by the following maps:
\begin{eqnarray}
\label{eq:Tw}
\lefteqn{
s_{ij}^{t+\frac{1}{2}}
=\sum_{k=i-1}^{i+1}\sum_{l=j-1}^{j+1}a_{ijkl}^{t}s_{kl}^{t}
}
\hphantom{\hspace{7.4cm}}
\nonumber\\
\lefteqn{
+\sum_{k=i-1}^{i+1}\sum_{l=j-1}^{j+1}a_{ijkl}^{t}
f\left(h_{kl}^{t}\right)g\left(h_{kl}^{t}\right)\Delta s_{kl}^{t},
}
\hphantom{\hspace{7.4cm}}
\\
\lefteqn{
c_{ij}^{t+\frac{1}{2}}=\sum_{k=i-1}^{i+1}\sum_{l=j-1}^{j+1}a_{ijkl}^{t}c_{kl}^{t},
}
\hphantom{\hspace{7.4cm}}
\\
\lefteqn{
\myvec{v}_{ij}^{t+\frac{1}{2}}=\sum_{k=i-1}^{i+1}\sum_{l=j-1}^{j+1}a_{ijkl}^{t}\myvec{w}_{kl}^{t},
}
\hphantom{\hspace{7.4cm}}
\end{eqnarray}
where $a_{ijkl}^{t}$ is the weight of allocation from lattice point $kl$ to lattice point $ij$ \cite{Nozawa},
\begin{eqnarray}
\lefteqn{a_{ijkl}^{t}=a\left(\myvec{r}_{ij},\tilde{\myvec{r}}_{kl}^{t}\right)=a\left(i,j,\tilde{k},\tilde{l}\right)}
\hphantom{\hspace{7.4cm}}
\nonumber\\
\lefteqn{
=\left(\delta_{i\lfloor \tilde{k}\rfloor}\delta_{j\lfloor \tilde{l}\rfloor}
+\delta_{i\lfloor \tilde{k}\rfloor +1}\delta_{j\lfloor \tilde{l}\rfloor}
+\delta_{i\lfloor \tilde{k}\rfloor +1}\delta_{j\lfloor \tilde{l}\rfloor +1}\right.
}
\hphantom{\hspace{7.4cm}}
\nonumber\\
\lefteqn{
\left.+\,\delta_{i\lfloor \tilde{k}\rfloor}\delta_{j\lfloor \tilde{l}\rfloor +1}\right)
\times\left(1-\left|\tilde{k}-i\right|\right)\left(1-\left|\tilde{l}-j\right|\right).
}
\hphantom{\hspace{7.4cm}}
\end{eqnarray}
It gives the fraction of emulsion carried from the cell at $kl$ to $ij$. The symbol $\lfloor\,\rfloor$ denotes the floor function.

\subsubsection{Coalescence procedure}
\label{coalescence procedure}

\begin{figure}[h]
\begin{center}
  \includegraphics[scale=0.5]{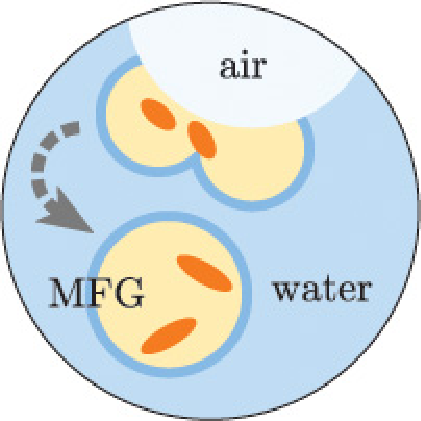}
\end{center}
\vspace{-0.5cm}
\caption{\footnotesize Conceptual illustration of the coalescence procedure $T_{c}$.
The MFGs partially coalesce through the aeration by whipping. The partial coalescence changes to complete coalescence due to a multiplicative effect between aeration and flocculation.
}
\label{fig:tc.eps}
\end{figure}

In the coalescence procedure $T_{c}$, the MFGs are first partially coalesced through the aeration due to whipping. After repeated whipping, a multiplicative effect between aeration and flocculation of the emulsion causes a rapid decrease in the surface activity of the MFG membranes, and the MFGs are finally completely coalesced \cite{Fujita}, as shown in Fig.\ref{fig:tc.eps}. This changes the surface energy of the emulsion to the cohesive energy of the emulsion.

The coalescence procedure $T_{c}$ is defined by the following maps using the occurrence factor $g(h_{ij}^{t})$ in Eq.\ref{eq:h(f)}:
\begin{eqnarray}
\label{eq:Tc}
\lefteqn{
s_{ij}^{t+1}=\gamma g\left(h_{ij}^{t+\frac{1}{2}}\right)s_{ij}^{t+\frac{1}{2}}
}
\hphantom{\hspace{7.4cm}}
\\
\lefteqn{
c_{ij}^{t+1}=c_{ij}^{t+\frac{1}{2}}+\gamma\left\{1-g\left(h_{ij}^{t+\frac{1}{2}}\right)\right\}s_{ij}^{t+\frac{1}{2}},
}
\hphantom{\hspace{7.4cm}}
\end{eqnarray}
where $\gamma$ is the coalescence coefficient which represents the conversion efficiency from surface energy $s_{ij}^{t}$ to cohesive energy $c_{ij}^{t}$. When $\gamma=1$, surface energy $s_{ij}^{t}$ is fully converted to cohesive energy $c_{ij}^{t}$, and thus emulsion energy $h_{ij}^{t}$ is conserved in the coalescence procedure $T_{c}$.

\subsubsection{Flocculation procedure}
\label{flocculation procedure}

\begin{figure}[h]
\begin{center}
  \includegraphics[scale=0.5]{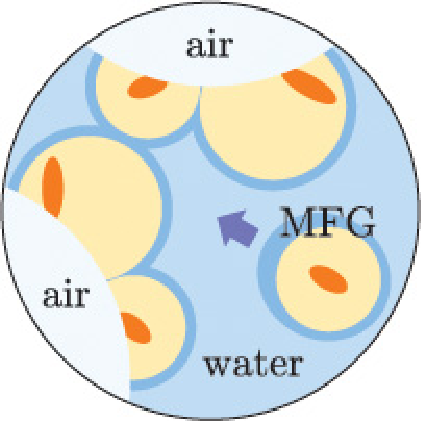}
\end{center}
\vspace{-0.5cm}
\caption{\footnotesize Conceptual illustration of the flocculation procedure $T_{f}$. The flow in the emulsion changes in the direction to demulsification due to the flocculation and Ostwald ripening of MFGs and air bubbles.}
\label{fig:tf.eps}
\end{figure}

In the flocculation procedure $T_{f}$, the flow of the emulsion is disturbed by the cohesive forces that drive flocculation and Ostwald ripening \cite{Fujita}, as shown in Fig.\ref{fig:tf.eps}. The cohesive forces increase the differences in the surface energy and cohesive energy (i.e., emulsion energy differences) between neighboring lattice points, causing air bubbles and MFGs to flocculate and grow. This induces the flow in a direction to demulsification that makes the emulsion state more unstable.

The flocculation procedure $T_{f}$ is defined by the following maps:
\begin{eqnarray}
\lefteqn{
v_{x\,ij}^{t+1}=v_{x\,ij}^{t+\frac{1}{2}}+\frac{\phi}{2}\left(f_{i+1j}^{t+1}-f_{i-1j}^{t+1}\right)
}
\hphantom{\hspace{7.4cm}}
\\
\lefteqn{
v_{y\,ij}^{t+1}=v_{y\,ij}^{t+\frac{1}{2}}+\frac{\phi}{2}\left(f_{ij+1}^{t+1}-f_{ij-1}^{t+1}\right),
}
\hphantom{\hspace{7.4cm}}
\end{eqnarray}
where $\phi$ is the flocculation coefficient, and $|\myvec{v}_{ij}^{t+1}|$ is assumed not to exceed one.

\subsection{Physical and chemical changes of the emulsion}
\label{physical and chemical changes of the emulsion}

\begin{figure*}[t]
\begin{center}
  \includegraphics[scale=0.425]{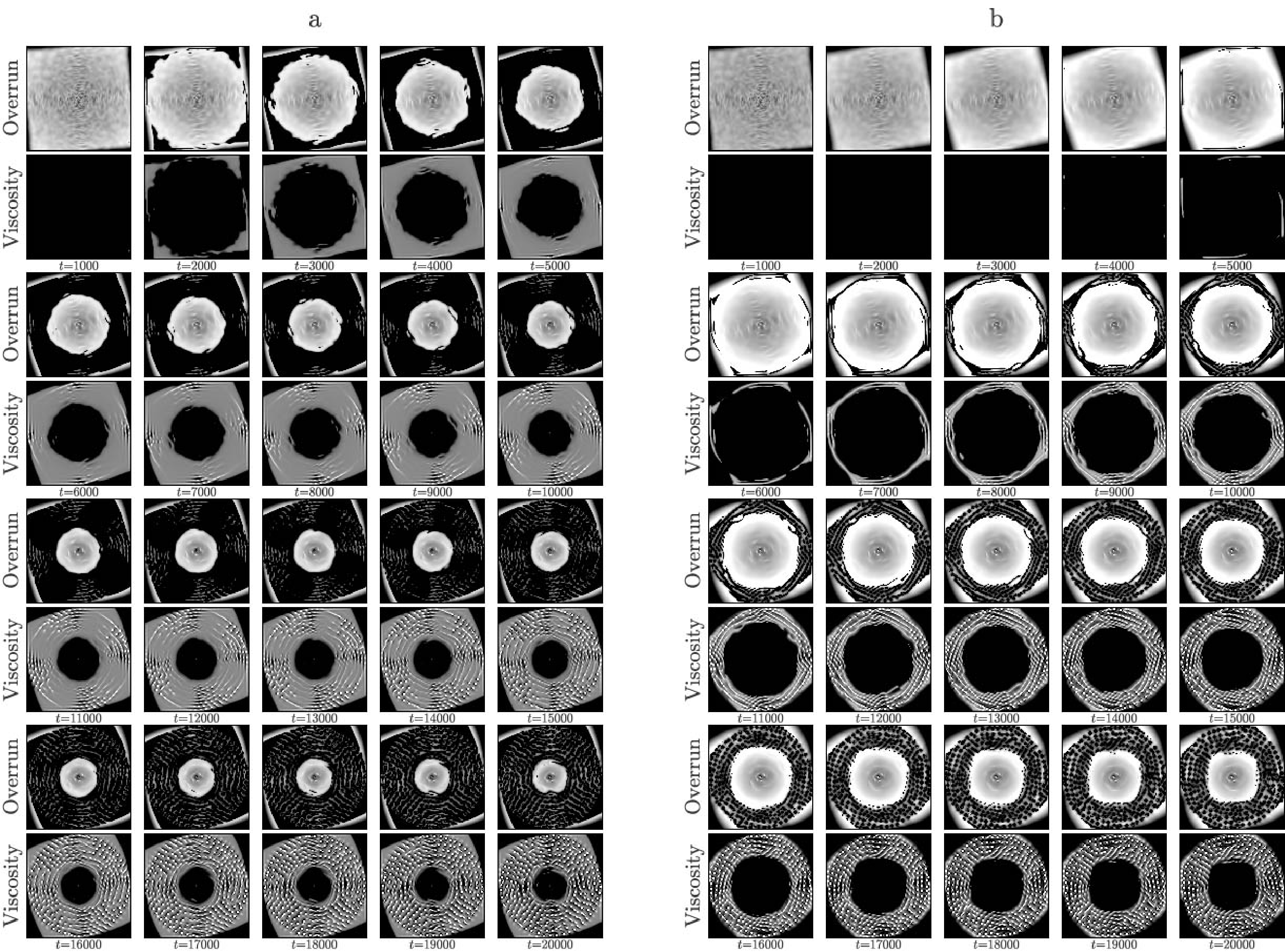}
\end{center}
\vspace{-0.5cm}
\caption{\footnotesize Snapshots of texture patterns at $t=1000,2000,\cdots,20000$. (a) At high whipping temperature. (b) At low whipping temperature. In the pair of snapshots at each time, the upper one labeled ``Overrun'' shows the logarithm of surface energy $s${\tiny $_{ij}^{t}$} and the lower one labeled ``Viscosity'' shows that of cohesive energy $c${\tiny $_{ij}^{t}$}, plotted in gray scale, respectively.}
\label{fig:ss_s_c_7_14.eps}
\vspace{-0.4cm}
\end{figure*}

The final construction step is to define the time evolution of the field variables describing the state change of the object by arranging a sequence of the procedures at each time step. In our model, we consider the order of physical and chemical changes of the emulsion in the phase inversion phenomenon and define the time evolution of surface energy $s_{ij}^{t}$, cohesive energy $c_{ij}^{t}$ and velocity $\myvec{v}_{ij}^{t}$ from discrete time $t$ to $t+1$ as follows:
\begin{eqnarray}
\lefteqn{
\left(
\begin{array}{lcl}
s_{ij}^{t}\rule{0pt}{1.2em} \\
c_{ij}^{t}\rule{0pt}{1.2em} \\
\myvec{v}_{ij}^{t}\rule{0pt}{1.2em} \\
\end{array}
\right)
\stackrel{\stackrel{whipping}{T_{w}}}{\longmapsto}
\left(
\begin{array}{lcl}
s_{ij}^{t+\frac{1}{2}}\rule{0pt}{1.2em} \\
c_{ij}^{t+\frac{1}{2}}\rule{0pt}{1.2em} \\
\myvec{v}_{ij}^{t+\frac{1}{2}}\rule{0pt}{1.2em} \\
\end{array}
\right)
}
\hphantom{\hspace{7.4cm}}
\nonumber\\
\lefteqn{
\stackrel{\stackrel{coalescence}{T_{c}}}{\longmapsto}
\left(
\begin{array}{lcl}
s_{ij}^{t+1}\rule{0pt}{1.2em} \\
c_{ij}^{t+1}\rule{0pt}{1.2em} \\
\myvec{v}_{ij}^{t+\frac{1}{2}}\rule{0pt}{1.2em} \\
\end{array}
\right)
\stackrel{\stackrel{flocculation}{T_{f}}}{\longmapsto}
\left(
\begin{array}{lcl}
s_{ij}^{t+1}\rule{0pt}{1.2em} \\
c_{ij}^{t+1}\rule{0pt}{1.2em} \\
\myvec{v}_{ij}^{t+1}\rule{0pt}{1.2em} \\
\end{array}
\right).
}
\hphantom{\hspace{7.4cm}}
\end{eqnarray}
Here we note that the superscript $t+\frac{1}{2}$ represents an intermediate time between discrete times $t$ and $t+1$ for convenience.

The simulations were performed with the following conditions and parameters:
Lattice size $N_{x}\times N_{y}$ is fixed as $100\times 100$;
at each lattice point $ij$ initial surface energy $s_{ij}^{0}$ is given by a uniform random number within $[0,0.2]$,
initial cohesive energy $c_{ij}^{0}$ is zero, and
initial velocity $\myvec{v}_{ij}^{0}$ is also zero;
the wall boundary conditions are assigned;
$\omega=0.01$, $\iota=100$, $\alpha=0.5$, $\tau=1$, $\kappa=2$, $\beta=30$, $\gamma=1$, and $\phi=0.01$.
Here we assume that the whipping temperature is determined by specifying the value of temperature-dependent threshold $\theta$ for the surface activity of MFG membranes: smaller values of $\theta$ correspond to higher temperatures \cite{Fujita,Dickinson,Walstra}.

\section{Simulation of two different phase inversion processes}
\label{simulation of two different phase inversion processes}

Two well-known phase inversion processes in dairy processing and confectionery \cite{Matsumoto,Okuyama,Ihara,Kajiwara} are reproduced by the proposed CML. One is a process at high whipping temperature where the low-aerated whipped cream phase-inverts via deaerated cream to soft butter, and the other is a process at low whipping temperature where the high-aerated whipped cream phase-inverts to hard butter.

Fig.\ref{fig:ss_s_c_7_14.eps} shows a series of snapshots of texture patterns at $t=1000,2000,\cdots,20000$.
It took only 160 seconds to obtain this result on a personal computer by using the fast computation of the proposed CML.
The snapshot labeled ``Overrun'' represents the aeration degree of the emulsion (whipped cream), with surface energy $s_{ij}^{t}$ plotted in gray scale, and the snapshot labeled ``Viscosity'' represents the flocculation and coalescence degree of the emulsion (butter), with cohesive energy $c_{ij}^{t}$ plotted in gray scale.

\begin{figure*}[t]
\begin{center}
  \includegraphics[scale=0.5]{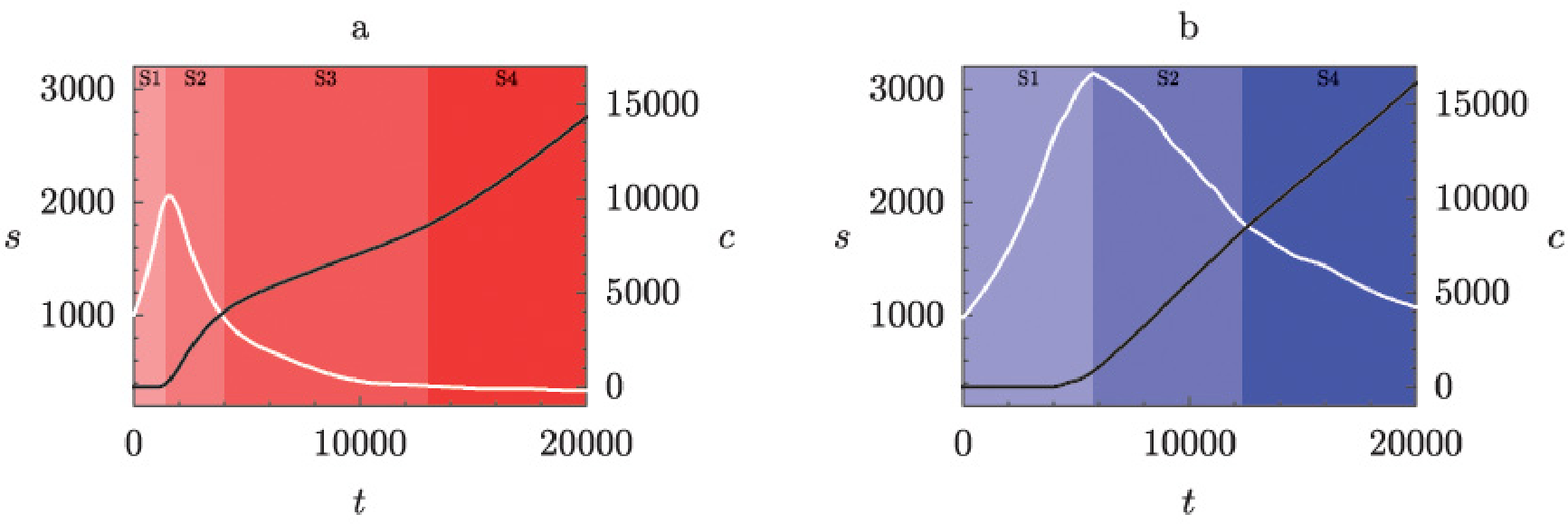}
\end{center}
\vspace{-0.5cm}
\caption{\footnotesize Overrun $s$ and viscosity $c$ curves with respect to whipping time $t$. (a) At high whipping temperature. (b) At low whipping temperature. The white curves represent the overrun changes, and the black curves the viscosity changes. The red color-coded background represents the four stages {\bf S1}-{\bf S4} at high whipping temperature, and the blue color-coded background the three stages {\bf S1}, {\bf S2}, {\bf S4}, respectively.}
\label{fig:sAt_t_cAt_t.eps}
\vspace{-0.4cm}
\end{figure*}

\subsection{Texture patterns at high whipping temperature}
\label{texture patterns at high whipping temperature}

We now make an overview of the simulated phase inversion process at high whipping temperature ($\theta=0.7$), shown in Fig.\ref{fig:ss_s_c_7_14.eps}a.
Overrun at $t=1000$ and $2000$: The emulsion is whipped, which leads to the partial coalescence of the MFGs, and thus it is aerated.
Overrun at $t=3000$ and $4000$: At high whipping temperature, the surface activity of MFG membranes is easy to reduce and the MFGs quickly coalesce, and thus the aeration of the emulsion remains low even when it is fully whipped, and then fades away.

Viscosity at $t=5000,\cdots,8000$: In the region where the aeration disappears and the deaerated emulsion appears, the weak flocculation of the coalesced MFGs begins to form streak patterns along whipping-induced flows, and their number and length gradually increase.
Viscosity at $t=9000$: In the outer region at high whipping speeds, the strong flocculation occurs with flow collisions, and very small newborn butter grains come into view.
Overrun and Viscosity at $t=10000,\cdots,15000$: The butter grains become larger while rapidly increasing their cohesive energy in the small reaeration derived from the strong flocculation, and grow here and there into small pea-sized clumps of butter.
Viscosity at $t=16000,\cdots,20000$: The grown butter grains disturb the whipping-induced flows, and the turbulent flows lead to further rapid growth of them. Thus, they fill the emulsion, and the phase inversion occurs.

\subsection{Texture patterns at low whipping temperature}
\label{texture patterns at low whipping temperature}

We now summarize features of the simulated phase inversion process at low whipping temperature ($\theta=1.4$), shown in Fig.\ref{fig:ss_s_c_7_14.eps}b.
Overrun at $t=1000$: The whipping results in the partial coalescence of the MFGs, and thus the aeration arises.
Overrun at $t=2000,\cdots,6000$: At low whipping temperature, the surface activity of MFG membranes is tough to reduce and the MFGs do not coalesce quickly, and thus a lot of air is introduced in with whipping, and then the aeration increases.

Overrun and Viscosity at $t=7000,\cdots,12000$: In the fully-whipped outer region, the coalescence of the MFGs finally begins, and a large amount of air held in accelerates the flocculation of the MFGs, and thus long streak patterns are formed one after another with fine intervals along whipping-induced flows.
Overrun and Viscosity at $t=13000,\cdots,t=20000$: The fine streak patterns cause collision of the flows and subsequent strong flocculation of the MFGs, and in the aeration, the small neighboring butter grains rapidly flocculate to form small pea-sized clumps of butter everywhere in the emulsion, and thus the phase inversion occurs.

\begin{figure*}[t]
\begin{center}
  \includegraphics[scale=0.5035]{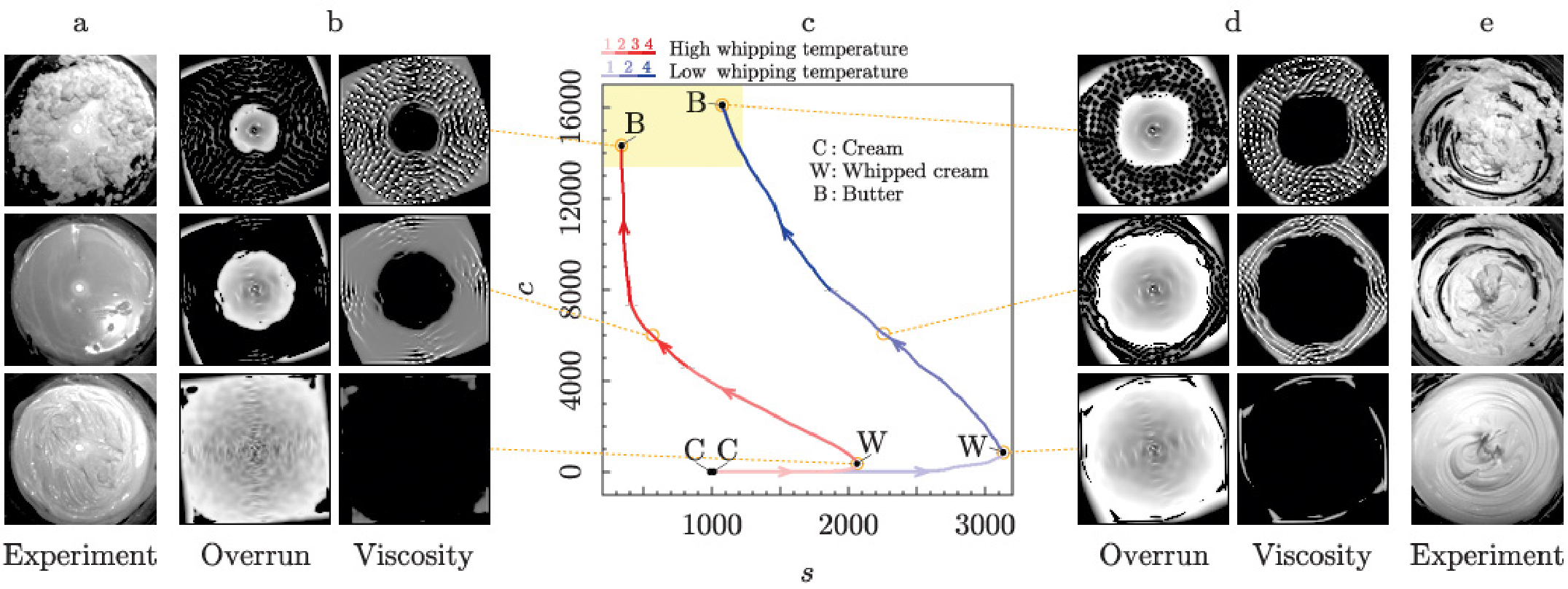}
\end{center}
\vspace{-0.5cm}
\caption{\footnotesize Two different phase inversion process curves at high and low whipping temperatures on the viscosity-overrun ($c$-$s$) plane. (a) Experimental images at high whipping temperature. (b) Texture patterns of overrun and viscosity at high whipping temperature. (c) $c$-$s$ plane. The red and blue curves are the emulsion state changes at high and low whipping temperatures, respectively. Each curve starts from point C to point B via point W, and is color-coded for each stage. The yellow rectangle represents a butter-formed area. (d) Texture patterns of overrun and viscosity at low whipping temperature. (e) Experimental images at low whipping temperature.}
\label{fig:cs_cur.eps}
\vspace{-0.4cm}
\end{figure*}

\section{Analysis of two different phase inversion processes}
\label{analysis of two different phase inversion processes}

As described in the previous section, the phase inversion processes are significantly different at high whipping temperatures and at low whipping temperatures.
Fig.\ref{fig:sAt_t_cAt_t.eps}a and Fig.\ref{fig:sAt_t_cAt_t.eps}b show the change in overrun $s$ (the white curves) and viscosity $c$ (the black curves) with respect to whipping time $t$ at high and low whipping temperatures, respectively.
Both results are at least qualitatively consistent with experimental results \cite{Matsumoto,Okuyama,Ihara} and confectionery techniques \cite{Kajiwara}.
Here, the overrun $s$ and the viscosity $c$ are the total surface energy and the total cohesive energy per one whipping respectively, given by
\begin{eqnarray}
\lefteqn{
s
=\frac{1}{\iota}\sum_{t'=t}^{\mathstrut t+\iota-1}\sum_{i=0}^{N_{\mathstrut x}-1}\sum_{j=0}^{N_{\mathstrut y}-1}s_{ij}^{t'},
}
\hphantom{\hspace{7.4cm}}
\\
\lefteqn{
c
=\frac{1}{\iota}\sum_{t'=t}^{\mathstrut t+\iota-1}\sum_{i=0}^{N_{\mathstrut x}-1}\sum_{j=0}^{N_{\mathstrut y}-1}c_{ij}^{t'}.
}
\hphantom{\hspace{7.4cm}}
\end{eqnarray}
Note that viscosity $c$ corresponds to the serum viscosity unaffected by overrun $s$, namely the viscosity of the deaerated emulsion in the experiments \cite{Noda,Ihara}.

Fig.\ref{fig:cs_cur.eps} shows the viscosity-overrun ($c$-$s$) plane (Fig.\ref{fig:cs_cur.eps}c), which is one of the state diagrams \cite{Levine}, with the texture patterns (Figs.\ref{fig:cs_cur.eps}b and \ref{fig:cs_cur.eps}d) and corresponding experimental images (Figs.\ref{fig:cs_cur.eps}a and \ref{fig:cs_cur.eps}e). The two different phase inversion processes are characterized on the $c$-$s$ plane by the red curve of high whipping temperature and the blue curve of low whipping temperature.

\subsection{Viscosity dominance at high whipping temperature}
\label{viscosity dominance at high whipping temperature}

At high whipping temperature, the increase in viscosity $c$ is dominant over the increase in overrun $s$ since the surface activity of MFG membranes is easy to reduce. We shall call this dynamic property a viscosity dominance in phase inversion.

The viscosity dominance is shown in Fig.\ref{fig:sAt_t_cAt_t.eps}a as the viscosity curve (the black curve) that quickly exceeds the overrun curve (the white curve).
The viscosity curve represents the four stages of change in viscosity $c$:
{\bf S1} is a slight increase with aeration; {\bf S2} is a rapid increase with coalescence; {\bf S3} is a gradual increase with weak flocculation; {\bf S4} is a steep increase with strong flocculation.
On the other hand, the overrun curve represents the increase and decrease stages of the change in overrun $s$: {\bf S1} is a small increase with aeration; {\bf S2}-{\bf S4} is an exponential decay with deaeration.
{\bf S3} is exhibited only in the viscosity dominance, since deaerated emulsion does not appear in the overrun dominance (as described in the next subsection).

The viscosity dominance is also shown on the $c$-$s$ plane in Fig.\ref{fig:cs_cur.eps}c as the process curve (the red curve) color-coded for each of the four stages {\bf S1}-{\bf S4}.
On the process curve, the emulsion state starts at point C with low $s$ and low $c$, i.e., fresh cream, and reaches point W with low-high $s$ and low $c$, i.e., low-aerated whipped cream, through {\bf S1} (the lightest red part of the curve).
Then, the state changes through the intermediate state with middle $s$ and low $c$ of {\bf S2} (the second lightest red part of the curve), that with low $s$ and low-middle $c$ of {\bf S3} (the third lightest red part of the curve), and that with low $s$ and high-middle $c$ of {\bf S4} (the red part of the curve) to point B with low $s$ and low-high $c$, i.e., soft butter.

Along the process curve, we focus on some distinctive texture patterns in Fig.\ref{fig:cs_cur.eps}b, and present the experimental images in Fig.\ref{fig:cs_cur.eps}a where the aeration, flocculation, and coalescence features are consistent with these texture patterns.
The experimental images were taken in a phase inversion process at the high whipping temperature of about \SI{17}{\degreeCelsius}.
The bottom row of Figs.\ref{fig:cs_cur.eps}b and \ref{fig:cs_cur.eps}a is the texture patterns and the experimental image of the low-aerated fully-whipped emulsion (whipped cream) located at the maximum overrun point W between {\bf S1} and {\bf S2}, respectively.
The middle row of Figs.\ref{fig:cs_cur.eps}b and \ref{fig:cs_cur.eps}a is those of the deaerated intermediate emulsion located at a point in {\bf S3}, respectively.
The top row of Figs.\ref{fig:cs_cur.eps}b and \ref{fig:cs_cur.eps}a is those of the coalesced phase-inverted emulsion (soft butter) located at point B in {\bf S4}, respectively.

\subsection{Overrun dominance at low whipping temperature}
\label{overrun dominance at low whipping temperature}

At low whipping temperature, the increase in overrun $s$ is dominant over the increase in viscosity $c$ since the surface activity of MFG membranes is tough to reduce. We shall call this dynamic property an overrun dominance in phase inversion.

The overrun dominance is shown in Fig.\ref{fig:sAt_t_cAt_t.eps}b as the overrun curve (the white curve) that is not quickly exceeded by the viscosity curve (the black curve).
The overrun curve represents the increase and decrease stages of the change in overrun $s$: {\bf S1} is a large increase with aeration; {\bf S2}, {\bf S4} is a power decay without deaeration.
On the other hand, the viscosity curve represents the three stages {\bf S1}, {\bf S2}, {\bf S4} of change in viscosity $c$.
{\bf S3} is not exhibited at all in the overrun dominance, since the aeration continues to remain and deaerated emulsion does not appear.

The overrun dominance is also shown on the $c$-$s$ plane in Fig.\ref{fig:cs_cur.eps}c as the process curve (the blue curve) color-coded for each of the three stages {\bf S1}, {\bf S2}, {\bf S4}.
On the process curve, the emulsion state starts at point C with low $s$ and low $c$, i.e., fresh cream, and reaches point W with high $s$ and low $c$, i.e., high-aerated whipped cream, through {\bf S1} (the lightest blue part of the curve).
Then, the state changes through the intermediate state with high $s$ and low-middle $c$ of {\bf S2} (the second lightest blue part of the curve), and that with middle $s$ and high-middle $c$ of {\bf S4} (the blue part of the curve) to point B with low-middle $s$ and high $c$, i.e., hard butter.

Along the process curve, we focus on some distinctive texture patterns in Fig.\ref{fig:cs_cur.eps}d, and present the experimental images in Fig.\ref{fig:cs_cur.eps}e where the aeration, flocculation, and coalescence features are consistent with these texture patterns.
The experimental images were taken in a phase inversion process at the low whipping temperature of about \SI{9}{\degreeCelsius}.
The bottom row of Figs.\ref{fig:cs_cur.eps}d and \ref{fig:cs_cur.eps}e is the texture patterns and the experimental image of the high-aerated fully-whipped emulsion (whipped cream) located at the maximum overrun point W between {\bf S1} and {\bf S2}, respectively.
The middle row of Figs.\ref{fig:cs_cur.eps}d and \ref{fig:cs_cur.eps}e is those of the aerated intermediate emulsion located at a point in {\bf S2}, respectively.
The top row of Figs.\ref{fig:cs_cur.eps}d and \ref{fig:cs_cur.eps}e is those of the flocculated phase-inverted emulsion (hard butter) located at point B in {\bf S4}, respectively.

\section{Summary and discussion}
\label{summary and discussion}

In this paper, we have proposed a minimal CML for reproducing phase inversion processes from fresh cream to whipped cream to butter caused by mechanical whipping.
In the proposed CML, the two-dimensional square lattice was used to introduce coordinates to an emulsion viewed from above, and the surface energy, cohesive energy, and velocity (flow) were assigned to the lattice as the physical and chemical field variables of the emulsion, and the whipping, coalescence, and flocculation procedures were formulated as a minimal set of procedures describing the elementary processes of physical and chemical changes of the emulsion.

In each procedure the following points were taken into account.
In the whipping procedure: a relaxation difference between water and MFGs in the whipping-induced flow of the emulsion; a membrane deformation at the mesoscale generated by the relaxation difference; the aeration at the macroscale led from the membrane deformation.
In the coalescence procedure: a rapid decrease in the surface activity of MFG membranes caused by a multiplicative effect between aeration and flocculation; the change from partial to complete coalescence of MFGs due to the rapid decrease in surface activity.
In the flocculation procedure: cohesive forces in flocculation and Ostwald ripening; a flow of the emulsion in the direction of demulsification induced by the cohesive forces.

The proposed CML has successfully simulated two well-known phase inversion processes in dairy processing and confectionery at high and low whipping temperatures.
For both phase inversion processes, the changes in overrun and viscosity of the simulated virtual emulsion with respect to whipping time are at least qualitatively consistent with the experimental results \cite{Matsumoto,Okuyama,Ihara} and the confectionery techniques \cite{Kajiwara}.
We have introduced a viscosity-overrun plane and characterized each process as it shows distinctive spatial patterns of overrun and viscosity.
At high whipping temperature, the viscosity dominance consisting of the four stages appears on the viscosity-overrun plane since the viscosity increase exceeds the overrun increase early;
whereas at low whipping temperature, the overrun dominance consisting of the three stages appears on the viscosity-overrun plane since the viscosity increase exceeds the overrun increase later.

The differences in the phase inversion processes between viscosity dominance and overrun dominance are expected to have a significant impact on the texture and quality design of the processed products, whipped cream and butter.
For example, in the butter-formed area on the viscosity-overrun plane (the yellow rectangle in Fig.\ref{fig:cs_cur.eps}c), the butter obtained in the process of viscosity dominance was of low overrun and viscosity (point B on the red curve in Fig.\ref{fig:cs_cur.eps}c) and had, so to say, ``creamy and soft'' texture (the top row of Figs.\ref{fig:cs_cur.eps}b and \ref{fig:cs_cur.eps}a).
On the other hand, the butter obtained in the process of overrun dominance was of high overrun and viscosity (point B on the blue curve in Fig.\ref{fig:cs_cur.eps}c) and had, so to say, ``fluffy and hard'' texture (the top row of Figs.\ref{fig:cs_cur.eps}d and \ref{fig:cs_cur.eps}e).
We will report elsewhere on a complex systems approach to new texture and quality design processes that induce the spontaneous formation or self-organization of processed products with target textures and qualities (e.g., butter with fluffy and creamy texture and moderate firmness) by searching for and controlling the processing parameters in the CML procedures such as whipping temperature and speed, and membrane properties.
Of course, the suggestive simulation results of CML supported by its fast computation will serve as a compass there, as discussed in Section \ref{introduction}.

In addition to the macroscopic perspectives such as overrun and viscosity, more microscopic perspectives such as air bubble sizes and butter grain sizes are also important for texture design and especially for quality design.
In future work, we intend to estimate the sizes of air bubbles and butter grains from overrun and viscosity, based on both interface and colloid science and statistical physics.

\section*{Acknowledgement}

The authors would like to thank Y. Masubuchi for valuable suggestions and for giving them the opportunity to perform phase inversion experiments.
E.N. also would like to express her sincere gratitude to p$\hat{\rm a}$tissier M. Kajiwara and his family, who gave her careful guidance on how to make whipped cream and butter, as well as providing her with a wealth of knowledge on confectionery.
E.N. was supported by The Japanese Dairy Science Association Foundation.

\section*{Author Declarations}

\subsection*{Conflict of Interest}

The authors have no conflicts to disclose.

\subsection*{Author Contributions}
{\bf Erika Nozawa}: Conceptualization (lead); Methodology (lead); Software (lead); Formal analysis (lead); Writing - original draft (lead); Writing - review and editing (equal); Funding acquisition (lead). {\bf Tetsuo Deguchi}: Conceptualization (supporting); Writing - original draft (supporting); Writing - review and editing (equal).

\section*{Data Availability}

The data that support the findings of this study are available from the corresponding author upon reasonable request.

\end{document}